\documentstyle[prl,epsfig,floats,aps, twocolumn]{revtex}
\begin{document}

\draft

\wideabs{

\title {Percolating through networks of random thresholds: 
      \linebreak
	Finite temperature electron tunneling in metal nanocrystal
      arrays}

\author{Raghuveer Parthasarathy$^{1,}$, Xiao-Min Lin$^{1,2}$,
Klara Elteto$^1$,
T. F. Rosenbaum$^1$, and Heinrich M. Jaeger$^1$}
\address{$^1$ James Franck Institute and Department of Physics,
  University of Chicago, Chicago, IL 60637}
\address{$^2$ Materials Science Division and Chemistry Division,
Argonne National Laboratory, Argonne, IL 60439}

\date{\today}

\maketitle

\begin{abstract}
We investigate how temperature affects transport through large networks of nonlinear conductances with distributed thresholds.  In monolayers of weakly-coupled gold nanocrystals, quenched charge disorder produces a range of local thresholds for the onset of electron tunneling. Our measurements delineate two regimes separated by a cross-over temperature $T^*$. Up to $T^*$ the nonlinear zero-temperature shape of the current-voltage curves survives, but with a threshold voltage for conduction that decreases linearly with temperature. Above $T^*$ the threshold vanishes and the low-bias conductance increases rapidly with temperature. We develop a model that accounts for these findings and predicts $T^*$.
\end {abstract}

}

Densely packed arrays of nanocrystals form a new class of ``artificial
solids''
with tunable electronic, magnetic and optical properties
\cite{black,doty01,leatherdale00,andres,sampaio2001,ginger00,prltransport}. 
These properties stem from single-electron charging and quantum confinement
energies on the individual particle level, mediated by the coupling to
neighboring particles \cite{collier98}. However, even for the simplest such
artificial solid, a close-packed monolayer of metallic nanoparticles, the
effect of temperature on the overall transport properties is poorly
understood.

At low temperatures, metal nanoparticle arrays show pronounced
nonlinear current-voltage ($IV$) characteristics, with a well-defined
threshold above which the applied bias has to be raised for conduction to
occur. Similar threshold behavior occurs in a large family of disordered
nonlinear systems, including the depinning of magnetic flux bundles in
type-II superconductors \cite{flux} and the onset of motion in charge-density
waves trapped by impurities\cite{CDW}. In metal nanoparticle arrays the
disorder arises, even for perfectly ordered particle arrangements, from
quenched background charges in the substrate or the matrix surrounding the
particles.  These quenched charges induce random shifts in the local
nanoparticle potentials and lead to a distribution of electrostatic energy
costs incurred by electrons as they tunnel from particle to particle. The
system behaves as a network of ultrasmall, Coulomb-blockade-type tunnel
junctions \cite{grabert} with a distribution of threshold energies.  For
$T=0$, a key result was obtained a decade ago by Middleton and Wingreen (MW)
\cite{mw} who showed through simulations and a scaling model that the
effective voltage threshold $V_t$ is proportional to the array length
and that the current-voltage ($IV$) characteristics exhibit power law scaling
of the total current, $I$, with excess voltage, $V - V_t$.  These predictions
are in line with experimental results on a wide range of nanocrystal arrays
\cite{black,andres,sampaio2001,prltransport,collier98}. More recently,
the effect of structural disorder on the scaling behavior was
investigated by experiments  and reproduced by simulations 
\cite{prltransport,reichhardt,ancona01}.

For finite temperatures, on the other hand, the literature to date shows no
consensus. In single-particle systems the shape of the $IV$ curves near
threshold is exponentially sensitive to temperature \cite{grabert}. For
arrays, Heath's group reported that increasing temperature rapidly rounds 
out any nonlinearities \cite{sampaio2001,collier98,heath02}. Black {\em et
al.} also observed linear behavior above 40K \cite{black}. Other recent work
found that finite temperature had a much
weaker effect on the shape of the $IV$ characteristics and on the threshold
$V_t$ \cite{prltransport,ancona01,bez,reichhardt03}.  Theoretical approaches that could calculate temperature-dependent
$IV$s for large arrays are currently not available. 

To address this issue, we have performed systematic experiments on large,
well-characterized two-dimensional arrays of gold nanocrystals.  Our results
show that the $IV$ curves retain their low-T, nonlinear scaling properties to
remarkably high temperatures and that $V_t$ decreases only  {\em linearly}
with T. We can explain these findings by assuming that there is a subset of
junctions, percolating across the array, for which thermal fluctuations
effectively have removed the Coulomb blockade. The starting point is the
picture developed by MW  of charge fronts propagating through the
array\cite{mw}. However, our model goes beyond MW in three key aspects:  it
introduces a way to deal with the effect of temperature on a
{\em distribution} of charging energies, it goes beyond square lattices and
applies to arbitrary regular networks, and it can explicitly take into
account capacitive coupling between neighboring islands, as is appropriate for
close-packed arrays. We believe that this very general approach may also be
applicable to other systems with distributions of local threshold values.

Our arrays consisted of monolayers of 1-dodecanethiol-ligated gold
nanocrystals, synthesized and deposited from solution onto silicon nitride
``window'' substrates  \cite{xmlnano,xmlarray}.
Several arrays were deposited onto substrates with thicker $\rm{Si}_3\rm{N}_4$
layers and no windows, more suitable for measurements above 200K. Eleven
arrays were studied ranging from monolayers with long-range hexagonal order,
with domains of typical size equal to the
electrode spacing, to monolayers with some disorder, including
voids or localized regions of double layers not more than 20\% in area
fraction. In each sample, the nanocrystals were monodisperse to within 5\% as 
checked by transmission electron microscopy (TEM).
Different samples had mean core diameters $4.5\rm{nm} < d <
6.5\rm{nm}$, and mean spacings $1.5\rm{nm} < s < 2.6\rm{nm}$. The relevant Coulomb charging
energy, $\Delta E$, is the change in electrostatic energy of the system due
to an electron tunneling from one particle to the next. $\Delta E$ depends on
the background charges on all particles involved. However, because of
compensation by mobile electrons, it suffices to consider effective
background charges within [-e/2, +e/2] and the distribution of charging
energies, $P(\Delta E)$, only extends to $\pm \Delta E_{max}$. For
close-packed monolayers, keeping only interactions between the two particles
forming the junction and the surrounding 8 nearest neighbors, to good
approximation $\Delta E_{max} = e^2({\bf C}^{-1}_{11} + 2{\bf C}^{-1}_{12}) =
e^2{\bf C}^{-1}_{11}(1 + 2\gamma)$, with $\gamma = {\bf C}^{-1}_{12}/{\bf
C}^{-1}_{11}$ where ${\bf C}^{-1}$ is the inverse capacitance matrix. From measurements of d and s, and taking a dielectric
constant $\epsilon = 2$ for the alkanethiol, the inverse capacitance matrix elements for such
10-particle system were calculated numerically  \cite{fastcap}. This gave 
${\bf C}^{-1}_{11}$ of order $10^{18} \rm{F}^{-1}$ and typical values for $\gamma \approx 0.36$. All
arrays had a fixed width ($2 \mu$m, or $\approx 270$ particles) and lengths,
$N$, ranging from 27 - 170 particles, set by the electrode separation. For
samples on window substrates, $N$ was measured by direct inspection with TEM
after the transport measurements were  completed. Otherwise, $N$ was
estimated from the electrode separation. Samples were cooled in vacuum, and
dc $IV$ curves were taken using Keithley 614 electrometers at voltage ramp
rates $5-25 \rm{mV/s}$.

At low temperatures, all arrays showed a clear voltage threshold for
conduction, indicating strong Coulomb blockade behavior.  Close inspection of
the $IV$ curves (see Fig.1) reveals that the differential conductance, $g
\equiv {\rm d}I/{\rm d}V$, for a given current $I$ is independent of
temperature below a sample-dependent characteristic temperature in the range
100 K - 200 K. The $IV$ curves keep their shape and simply shift to lower
voltages with increasing temperature. More quantitatively, we find that the
$IV$s can be collapsed on top of each other by translating them by an amount
$V_{\rm shift}(T) < 0$. Defining an
{\em effective} threshold $V_{\rm t}(T) = V_{\rm t}(0) + V_{\rm shift}(T)$,
we
can collapse all traces onto a master curve as shown in Fig.1b. $|V_{\rm
shift}(T^*)| = V_{\rm t} (0)$ defines the temperature, $T^*$, at which the
array threshold first reaches zero. Once $T > T^*$, the $IV$s exhibit
non-zero differential conductance at zero bias (Fig.1a inset) prohibiting a
data collapse at small bias.  However, at higher bias the $IV$s still
collapse when translated by an amount $|V_{\rm shift}(T)| > V_{\rm t}(0)$. 
The regime above $T^*$ thus is characterized by a negative {\em effective}
threshold and low-bias data falling below the master curve.

\begin{figure}[tb]
\begin{center}
\includegraphics[width=8.6cm]{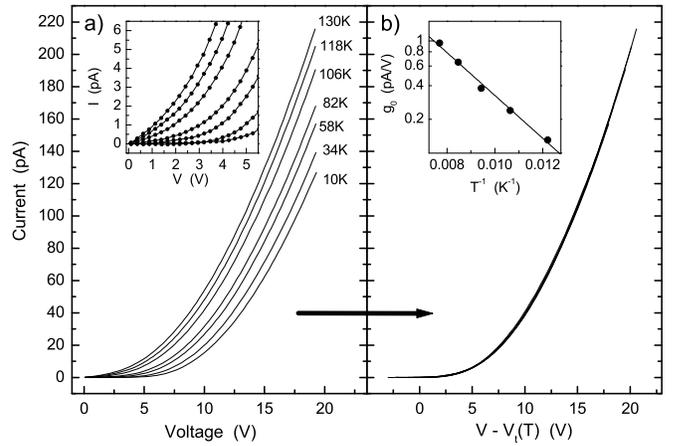}
\end{center}
\caption{a) Evolution of $IV$ curves with temperature for an array of length
$N = 128$. The $IV$s are fully symmetric with respect to bias voltage
reversal. Above $T^*$ the $IV$s no longer have a finite voltage threshold, as
shown in the inset.
b) Below $T^*$, $IV$s at different temperatures can be collapsed by
translating the voltage scale by an amount $V_{\rm shift}(T)$, with no
alteration of the current scale. The inset shows the zero-bias conductance,
$g_0 = {\rm d}I/{\rm d}V$, as a function of temperature.
}
\label{Fig1}
\end{figure}

For highly ordered arrays (bottom curve in Fig.2a) MW's $T=0$ scaling law,
$I \sim (V - V_t)^\zeta$, remains valid up to $T^*$ with
temperature-independent exponent $\zeta = 2.25 \pm 0.1$. Details of the
functional form of $I(V)$ in different arrays affect the value of $T^*$ but
are irrelevant as far as the collapse is concerned. Even for slightly
disordered arrays, characterized by a somewhat {\em s}-shaped $IV$
\cite{prltransport} in Fig.2, collapse can be achieved. For all arrays
studied, $V_{\rm t}(T)$, the only adjustable parameter in the scaling
procedure, drops essentially {\em linearly} with increasing temperature
(Fig.2b). Both $V_{\rm t}(0)$ and the slope of this depression increase with
array size $N$. Furthermore, for similar $V_t(0)$ the slope decreases with
increasing charging energy.

\begin{figure}[tb]
\begin{center}
\includegraphics[width=8.6cm]{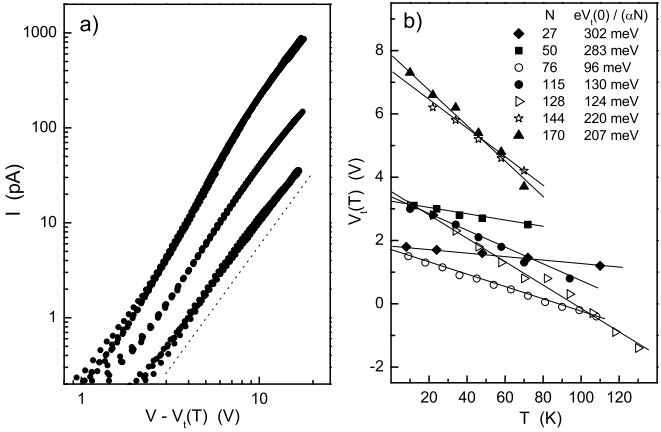}
\end{center}
\caption{a) Scaling behavior of current $I$ with excess voltage $V -
V_{t}(T)$ above threshold for same array as in Fig.1 (middle curve) and two
other arrays with different degrees of disorder. For each array, 4 different 
temperatures are plotted. The dotted line corresponds
to $\zeta = 2.27$.
b) The effective threshold as a function of temperature for several arrays of length $N$ and effective charging energy $eV_t(0) / \alpha N$.}
\label{Fig2}
\end{figure}

This slow, linear depression of the voltage threshold with temperature is
surprising and unexpected from the theory of few-junction networks
\cite{grabert} or from other nonlinear systems such as charge-density waves
\cite{CDW}. A similar trend was first noted in 1D chains of carbon
nanospheres by Bezryadin {\em et al.} \cite{bez}, and by Ancona {\em et al.}
\cite{ancona01} in small 2D arrays. Bezryadin {\em et al.} proposed that
thermal fluctuations reduce the charging energy by $k_BT$ in all particles and
that therefore $V_{t}(T)$ should decrease as $Nk_BT/e$. However, this simple
picture cannot account for the slope values in Fig. 2b. We now outline a
model that captures the linear dependence of $V_{t}(T)$, predicts the slope,
and, under appropriate rescaling, collapses the data in Fig.2 onto a
universal curve.

MW showed that the minimum bias voltage for charges to move all the way from
one electrode to the other depends on the {\em number}
of ``up-steps" along any given path. Each local up-step in potential, no
matter how small, requires an increment $\Delta V$ in the externally applied
bias for the charge front to advance, and thus contributes equally. 
($\Delta V=\Delta E_{max}/e$ for uncoupled particles or junctions surrounded by charge neutral neighbors; for the more general case see below.)
The
array threshold is set by the optimal path with the smallest total number of
up-steps and scales as $V_t = \alpha N \Delta V$ in both 1D and 2D \cite{mw}.
In 2D, $\alpha = 0.338$ for square lattices \cite{mw} and $\alpha = 0.227$
for triangular lattices \cite{longpaper}.  Down-steps do not gain energy
because all tunneling between particles is inelastic (no resonant tunneling
between sharp levels).  Hence, as long as the number of up-steps along a path
remains fixed, a gradual reduction in the associated energy step size,
induced by thermal fluctuations, will not reduce $V_{\rm t}$. We now argue
that $V_{\rm t}$ can be reduced if temperature eliminates some of the
up-steps completely. This occurs for the subset of junctions with step
magnitudes $|\Delta E|$ so small that they are washed out by thermal
fluctuations.  
 
We consider a scenario in which temperature has removed the nonlinear,
threshold-type Coulomb blockade behavior of a fraction $p(T)$ of the
junctions
(irrespective of the sign of the step), effectively turning them into linear
resistors. The effect of temperature, then, is a shift in threshold
proportional to $p(T)$.
At some temperature, $T^*$, the fraction $p(T^*)$ will become large enough
for a continuous path of junctions without discernable Coulomb blockade to
form a percolating sub-network that bridges the array.  At this point the
overall array voltage threshold will vanish. We therefore identify $p(T^*)$
with the percolation threshold, $p_c$, for the underlying lattice. Hence we
expect $V_{\rm t}(T) = V_{\rm t}(0) (1 - p(T)/p_c)$. For a triangular lattice
$p_c \approx 0.347$. To make explicit the temperature
dependence of the threshold voltage, we need to find $p(T)$. 
For a given distribution, $P(\Delta E)$, junctions corresponding to $|\Delta E |< b k_{\rm B} T$ are taken as 
effectively linearized. The factor $b$ accounts for
the broadening of the Fermi-Dirac distributions in the nanocrystals.
Electrons in the high-energy tail of one nanocrystal's distribution can
tunnel into the low-energy tail of accessible states of a neighboring
particle, reducing the energy cost by roughly $2k_{\rm B}T$. More accurately,
equating the reduction with the difference between the mean energies in the
high- and low-energy tails, we calculate $b = 2.4$ \cite{longpaper}. At any
given temperature, $p(T)$ is then given by the integral over $P(\Delta E)$
between the limits $\pm b k_B T$.

Consider first the case of two particles with random charge offsets in the range $[-e/2,e/2]$, forming
a representative tunnel junction, surrounded by 8 {\em neutral} neighbors. In
this case, $\Delta E_{max} = e^2{\bf C}^{-1}_{11}(1 -\gamma)$, and $P(\Delta
E)$ is a triangle with y-intercept $P(0) = 1/{\Delta E_{max}}$ (inset to Fig.3) \cite{longpaper}. Integrating $P(\Delta E)$ over the shaded area leads to
$p(T) = bx(2 - bx)$ with $x = k_B T P(0)$. Using $p(T^*) =
p_c$ and $b = 2.4$ we find $k_B T^* \approx 0.08/P(0)$. Note that $p(T)$ is
proportional to $T$ and $P(0)$ as long as $x<<1$, which is satisfied up to
temperatures far above $T^*$. Allowing for random offset charges also on the
8 neighbors rounds the corners and steepens the slope of $P(\Delta E)$. Still,
the overall triangular shape remains, $p(T) = 2bx + \mathcal{O}$$(x^3)$ and in the above 
considerations only the
numerical value for $P(0)$ needs to be modified. Now $\Delta
E_{max} = e^2{\bf C}^{-1}_{11}(1 + 2\gamma)$ and $P(0) = (e^2{\bf
C}^{-1}_{11})^{-1}(1 - 1.57 \gamma)/(1 - \gamma)^2$ \cite{longpaper}. 
For all arrays, independent of N, the model then predicts that the normalized threshold 
$V_{\rm t}(T) / V_{\rm t}(0)$ decreases linearly as a function of the
dimensionless variable $x = k_B T P(0)$, with a slope
$-2b/p_c$, crossing zero near $x = 0.072$. 
Within this picture, $1/P(0)$ can be thought of
as an effective charging energy per up-step that captures all effects of the local
packing geometry and the quenched charge distribution.

\begin{figure}[tb]
\begin{center}
\includegraphics[width=8.6cm]{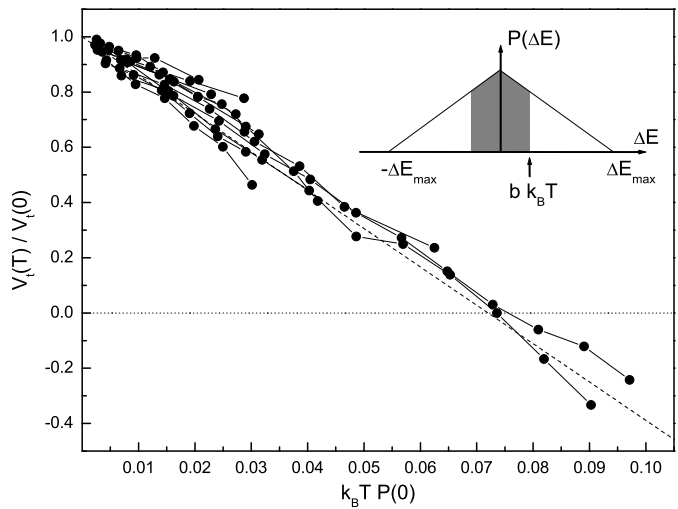}
\end{center}
\caption{Normalized threshold, $V_t(T)/V_t(0)$ as a function of dimensionless
temperature variable $x=k_B T P(0)$ for all arrays.
The dashed line is the model prediction.
Inset: Probability density of charging energies $P(\Delta E)$. With capacitive coupling in the presence of offset charges on all nearest neighbors, $P(\Delta E)$ evolves to a more rounded shape.}
\label{Fig3}
\end{figure}

These predictions agree well with our data (Fig.3) \cite{note}. To obtain $V_{\rm t}(0)$
we extrapolated $V_{\rm t}(T)$ back to $T = 0$ for each sample
(Fig.2b). To determine $P(0)$ two methods were used, giving independent
checks of the model. Equating $e\Delta V = 1/P(0)$ for the energy per up-step we obtained $P(0)$ from $V_{\rm t}(0) = \alpha N \Delta V$ using $N$ from the electrode separations and $\alpha = 0.227$.  In addition, for the samples where TEM images of sufficient
resolution were available to extract d and s, we were able to cross-check the values for $P(0)$ with those derived from direct numerical calculation of the ${\bf C}^{-1}_{ij}$ \cite{fastcap}.

The broad distribution of local energy differences $\Delta E$ guarantees
that, even if a fraction of steps is effectively removed, the underlying
concept of current paths meandering through a random potential landscape -
the premise of MW's scaling picture - remains valid. The general shape of the
$IV$s above $V_{\rm t}(T)$, which reflects how additional current paths open
up with increasing bias beyond threshold, therefore should remain unchanged
up to $T^*$. This is borne out by our data. Rounding of the Coulomb blockade
in the local junction $IV$ characteristics affects the quality of the data
collapse only at the smallest $V-V_{\rm t}(T)$ (Fig.2a).

The behavior for $T > T^*$ is quite different. Here the low-bias conductance
is dominated by a few junctions from the ($p_c/2$)-th percentile of $P(\Delta
E)$, since they are the key, "bottleneck" bonds in the lattice during
percolation at $T^*$.  Therefore, the low-bias $IV$s for the percolating path
resemble those of a single island system with washed-out Coulomb blockade,
for which the linear zero-bias conductance (Fig.1a inset) exhibits thermally
activated behavior \cite{grabert}: $g_0 (T) \sim \exp \left ( -U / k_B T
\right )$.
Our model predicts $U = b k_B T^* = 0.17 \Delta E_{max}$, the largest $\Delta E$ for the
percolating path. This $g_0$ will dominate the low-bias array conductance as
$T$ increases and additional, parallel paths open up that have larger $U$.
Because of these additional paths, we expect the measured effective
activation energy to be somewhat larger than $b k_B T^*$. Our data indeed are
compatible with simple activated behavior and a magnitude of $U$ 1 to 2 times
$b k_B T^*$. For example, for the array in Fig.1b inset $\Delta E_{max} = 124$ meV,
giving $b k_B T^* = 21$ meV, to be compared with $U \approx 38$ meV from the
slope of ${\rm ln}(g_0)$ vs $T^{-1}$. However, because of the limited (high)
temperature range available, our data cannot rule out other functional forms,
such as variable range hopping conduction as seen in Ref.\cite {heath02}.

Our results support a picture of thermally assisted quantum tunneling between
neighboring gold nanocrystal cores through the dodecanethiol ligand. Within
this picture, the ligands serve as a mechanical spacer between the particles
and they set the tunneling barrier height.  They do not introduce additional
states into the barrier, nor is the barrier low enough to be thermally hopped
over \cite{harris}, both of which may be the case with more complex ligands
\cite{andres}. We find no evidence of a regime with positive temperature
coefficient of resistance as seen in Ref. \cite{sampaio2001}. Such behavior
may require a stronger interparticle coupling and junction resistances much
closer to the quantum resistance $R_q = h/e^2$. In the weak coupling limit,
as in our arrays, the value of $T^*$ as well as the collapse of the
temperature-dependent threshold data onto a universal curve emerge naturally
from considering the interplay of tunneling and quenched charge disorder.

We thank Qiti Guo and Benjamin Lauderdale for experimental assistance,
and Eduard Antonyan, Terry Bigioni, Ilya Gruzberg, Alan Middleton, Toan Nguyen and Tom Witten for stimulating
discussions. XML acknowledges support
from DOE W-31-109-ENG-38. This work was supported by the UC-ANL Consortium
for Nanoscience Research and by the NSF MRSEC program under DMR-0213745.


\vspace{-0.2in} \references
\vspace{-0.5in}

\bibitem {black} C. T. Black, C. B. Murray, R. L. Sandstrom, and
S. Sun, Science {\bf 290}, 1131 (2000).

\bibitem {doty01} R. C. Doty, H. Yu, C. K. Shih, and B. Korgel,
J. Phys. Chem. B {\bf 105}, 8291 (2001).

\bibitem {leatherdale00} C. A. Leatherdale {\em et al.}, Phys. Rev. B
{\bf 62}, 2669 (2000).

\bibitem {andres} R. P. Andres {\em et al.}, Science {\bf 273}, 1690 (1996).

\bibitem {sampaio2001} J. F. Sampaio, K. C. Beverly, and J. R. Heath,
J. Phys. Chem. B {\bf 105}, 8797 (2001).

\bibitem {ginger00} D. S. Ginger and N. C. Greenham, J. Appl. Phys.
{\bf 87}, 1361 (2000).

\bibitem {prltransport} R. Parthasarathy, X.-M. Lin, and H. M. Jaeger,
Phys. Rev. Lett. {\bf 87}, 186807 (2001).

\bibitem {collier98} C. P. Collier, T. Vossmeyer, and J. R. Heath,
Annu. Rev. Phys. Chem. {\bf 49}, 371 (1998).

\bibitem {flux} S. Bhattacharya and M. J. Higgins, Phys. Rev. Lett. {\bf 70},
2617 (1993).

\bibitem {CDW} G. Gr\"uner , Rev. Mod. Phys. {\bf 60}, 1129 (1988).

\bibitem {grabert} {\em Single Charge Tunneling}, ed. H. Grabert and M. H.
Devoret (Plenum, New York, 1992).

\bibitem {mw} A. A. Middleton and N. S. Wingreen, Phys. Rev. Lett.
{\bf 71}, 3198 (1993).

\bibitem {reichhardt} C. Reichhardt and C. J. Olson Reichhardt, Phys. Rev.
Lett. {\bf 90}, 046802 (2003).

\bibitem {ancona01} M. G. Ancona {\em et al.}, Phys. Rev. B
{\bf 64}, 033408 (2001).

\bibitem {heath02} K. C. Beverly, J. F. Sampaio, and J. R. Heath, J. Phys.
Chem. B {\bf 106}, 2131 (2002).

\bibitem {bez} A. Bezryadin {\em et al.}, Appl. Phys. Lett.
{\bf 74}, 2699 (1999).

\bibitem {reichhardt03} C. Reichhardt and C.J. Olson Reichhardt, Phys. Rev. B {\bf 68}, 165305 (2003).

\bibitem {xmlnano} X.-M. Lin, {\em et al.}, J. Nanopart. Res. {\bf 2}, 157 (2000).

\bibitem {xmlarray} X.-M. Lin, H. M. Jaeger, C. M. Sorensen,
and K. J. Klabunde, J. Phys. Chem. B {\bf 105}, 3353 (2001).

\bibitem {fastcap} \textsc{fastcap} capacitance calculator program, MIT \copyright  1992. 

\bibitem {longpaper} For details about these calculations as well as supporting
simulations see K. Elteto {\em et al.} (to be published).

\bibitem {note} In 1D $p_c = 1$ and $\alpha = 1/2$, and our model predicts
$V_{t}(T) = V_{t}(0)(1-p(T)) \approx V_{t}(0) - 1.2Nk_{B}T/e$, reproducing
the behavior observed in \cite{bez} for nanoparticle chains.

\bibitem {harris} R. L. Lingle {\em et al.}, Chem. Phys. {\bf 205}, 191
(1996);
erratum {\em ibid.} {\bf 208}, 297 (1996).
\end {document}